\relax
\documentclass[letterpaper]{article} 
\usepackage[preprint] {nips_2018}  
\usepackage{times}  
\usepackage{helvet}  
\usepackage{courier}  
\usepackage{url}  
\usepackage{graphicx}  
\usepackage{enumitem}

\usepackage{amsmath,amsfonts,amssymb,amsthm}
\usepackage{commath}
\usepackage{bm}
\usepackage{booktabs}
\usepackage{nicefrac}
\usepackage{todonotes}
\usepackage[ruled]{algorithm2e}

\newcommand{\x}{x}
\newcommand{\z}{\vec{z}}
\newcommand\abbr[1]{\textsc{#1}}

\newcommand{\LL}{\mathcal{L}}
\newcommand{\E}{\mathbb{E}}

\DeclareRobustCommand{\uvec}[1]{{%
  \ifcsname uvec#1\endcsname
     \csname uvec#1\endcsname
   \else
    \bm{\hat{\mathbf{#1}}}%
   \fi
}}
\usepackage{authblk}
\usepackage{multirow}

\begin{document}
%
\title{PepCVAE: Semi-Supervised Targeted Design of Antimicrobial Peptide Molecules}

\author[ 1,3]{\small Payel Das\thanks{The first two authors contributed equally} }
\author[*1]{\small Kahini Wadhawan}
\author[2]{\small Oscar Chang\thanks{Work done as an IBM Research intern} }
\author[1]{\small Tom Sercu}
\author[1]{\small Cicero dos Santos}
\author[1]{\small Matthew Riemer}
\author[1]{\small Vijil Chenthamarakshan}
\author[1]{\small Inkit Padhi}
\author[1]{\small Aleksandra Mojsilovic}

\affil[1]{\footnotesize IBM Research AI, Yorktown Heights, NY 10598 }
\affil[2]{\footnotesize Data Science Institute, Columbia University, NY 10027}
\affil[3]{\footnotesize Applied Physics and Applied Mathematics Depeartment, Columbia University, NY 10027}
\affil[4]{\footnotesize {\{daspa,tsercu,cicerons,mdriemer,
ecvijil,aleksand\}@us.ibm.com} {\{kahini.wadhawan,inkit.padhi\}@ibm.com} oscar.chang@columbia.edu }

\maketitle 
\begin{abstract}
Given the emerging global threat of antimicrobial resistance, new methods for next-generation antimicrobial design are urgently needed. We report a peptide generation framework PepCVAE, based on a semi-supervised variational autoencoder (VAE) model, for designing novel antimicrobial peptide (AMP) sequences. Our model learns a rich latent space of the biological peptide context by taking advantage of abundant, unlabeled peptide sequences. The model further learns a disentangled antimicrobial attribute space by using the feedback from a jointly trained AMP classifier that uses limited labeled instances. The disentangled representation allows for controllable generation of AMPs. Extensive analysis of the PepCVAE-generated sequences reveals superior performance of our model in comparison to a plain VAE, as PepCVAE  generates novel AMP sequences with higher long-range diversity, while being closer to the training distribution of biological peptides. 
\end{abstract}

\section{Introduction}
\noindent  Hospital-acquired infection is a serious global health concern and is the sixth leading cause of death in the United States, with an estimated cost of ~\$10 billion annually \citep{peleg2010hospital}. 60-70\% of hospital-acquired infection is attributed to Gram-negative bacteria. 
Those bacteria are also efficient in creating antibiotic-resistant mutants. Each year, 30 million sepsis cases are reported worldwide, and potentially 5 million deaths occur as a result of antibiotic-resistant infections \citep{fleischmann2016assessment}. The estimated annual number of deaths due to direct antibacterial resistance (AMR) is reported to be at least 23,000 in US alone \citep{cdcwebsite} and 700,000 globally.
The emergence of multidrug-resistant bacterial strains, \textit{aka} priority pathogens \citep{cdcwebsite}, combined with the dry drug pipeline, advocates for urgent development of new approaches to fight AMR.
Antimicrobial peptides (AMP) or host defense peptides are peptide sequences typically  comprised of 10-50 amino acids. AMPs directly disrupt the bacterial membrane integrity, leading to membrane pore formation and membranolysis,
which is referred as antimicrobial activity. These peptides are found among all classes of life and  are considered as potential candidates for next-generation antibiotics, because of their  natural antimicrobial properties and a low propensity for development of resistance by microorganisms. 

Until now, discovery of many therapeutic molecules either happened by chance, \textit{e.g.} discovery of penicillin, or \textit{via} exhaustive combinatorial search \citep{porto2018silico}.
\textit{In cerebro} design of AMPs involves synthesis or modification of peptide sequences toward certain desired characteristics, \textit{e.g.} enhancing net positive charge and hydrophobicity, which favors interaction with negatively charged bacterial membrane.
Such approaches suffer from three main obstacles: (1) it is practically impossible to perform an exhaustive search and characterization of the original sequence space. (2) Hand-engineering and/or selecting features is frequently needed. 
And (3) it is often not possible to have control over the generation process during a trial-and-error method or brute-force search. Therefore, inverse  design of therapeutic molecules remains challenging. 

Recently, the combination of big datasets with machine learning methods, like deep generative models, has opened the door towards accelerated molecule discovery by using data-driven approaches. 
In fact, in recent years, popular deep generative models, such as generative adversarial networks (GAN) \citep{goodfellow2014generative} and variational autoencoders (VAE) \citep{kingma2013auto} have been successfully adapted to the development of new molecules \citep{kadurin2017drugan,blaschke2018application,gomez2018automatic,kusner2017grammar, jin2018junction}. 
Often, the molecule generation task is approached by formulating the design problem as a natural language generation problem, in which molecules are represented as SMILES strings, \textit{i.e.} sequences of characters. 
Similarly, biological molecule (peptide, nucleic acid) generation can be tackled by presenting the sequence as a text string of building blocks: 
\textit{e.g.} peptide as a string of 20 basic amino acid characters, and 
nucleic acid as a string of 4 basic nucleotide characters. It has been suggested that biological sequences exhibit characteristics typical of natural-language texts, such as ``signature-style'' word usage indicative of authors or topics. For example, at an unigram level, AMPs are reported to be rich in cationic (K, R) and hydrophobic (A, C, L) amino acids. The natural language processing and generation algorithms may therefore be adapted to ``biological language modeling''  \citep{osmanbeyoglu2011n,muller2018recurrent,nagarajan2018computational}. 
Therefore, it is safe to state. that by exploiting the advent of large-scale data from next-generation sequencing, generative machine learning algorithms can potentially accelerate the discovery process of novel AMPs.

The generative modeling task is difficult mostly because it needs to fulfill three main desiderata:
(1) \textbf{discrete sequence} data generation, 
which is more difficult than continuous data generation;
(2) \textbf{controlled} generation with certain attributes (\textit{e.g.} AMP characteristics) disentangled into controllable ``knobs'';
and (3) \textbf{diversity} of generated sequences.
The diversity of generated peptide sequences is one of the most important features that a good AMP generator should possess.
One primary reason for the difficulty on generating diverse sequences is that the generative models are usually trained on labeled data only
, which is relatively scarce and sparse and labeling at massive scale remains expensive.
Generated sequences from those models demonstrate high identity with known sequences in public databases and contain a restrictive set of amino acid patterns (see \citep{porto2018silico} and references therein).
Therefore, to achieve diversity, it is necessary to develop semi-supervised generative models that can simultaneously learn from the large unlabeled peptide sequence databases and a limited number of labeled sequences.


In this work, we employ a combination of VAE and an AMP classifier to learn the disentangled latent space of peptide sequences following \citep{hu2017toward},
and generate novel antimicrobial molecules by sampling from the latent space. We refer to this framework as PepCVAE. 
To this end, we collected and curated a new dataset consisting of two main parts:
(1) a large unlabeled dataset (1.6M samples) of peptide sequences, and (2) a smaller set (15k samples) of peptides labeled for antimicrobial activity/inactivity.
We demonstrate the advantage of our PepCVAE architecture and approach
by rigorously comparing the generated sequences with those from a simple VAE architecture trained only on AMP sequences. 
The results show that our semi-supervised VAE setup produces a more diverse set of biologically relevant AMP sequences. The proposed approach can therefore be applied to the general task of targeted design of novel molecules and materials. 


\section{Related Work}
\textbf{Sequence Generation:}
The simplest model one can consider for sequence generation involves a single recurrent neural network (RNN) language model 
that predicts most probable next token, given previous tokens. 
\citep{muller2018recurrent} have proposed a RNN-based peptide generator that is trained on a set of known AMP sequences.
A more versatile approach to sequence generation is based on the VAE framework,
which allows to sample new sequences based on a continuous latent space $z$.
\citeauthor{bowman2015generating} first used a VAE for probabilistic generation of natural language, which has more recently been adapted to molecular SMILES sequence generation \citep{gomez2018automatic}.
Sequence generation using GANs has also been the focus of recent research, although 
they require special techniques to deal with the non-differentiable nature of sequences
\citep{yu2017seqgan,kusner2016gans}.

\textbf{Controlled Sequence Generation:}
the controlled generation of text based on stylistic attributes has been the focus of a number of recent papers.
\citep{hu2017toward} have demonstrated that, in a variational encoder-decoder setup, feeding the generator with the attribute information along with the latent variable enables generation with control. 
Additionally,
\citeauthor{hu2017toward}'s method allows for semi-supervised learning.



\section{Methods}

\subsection{VAE-based Sequence Generation}
Variational  Autoencoder (VAE) is a class of generative models that build on the autoencoder (AE) framework by adding a particular type of regularization. 
VAE's regularization imposes a prior distribution $p(\z)$ to the latent codes $\z$.
This allows the sampling of new instances (not present in the training set) by sampling from the prior distribution $p(\z)$.
Concretely,
the encoder in VAE parametrizes an approximate posterior distribution over $\z$ with a neural network conditioned on the input $x$.
The prior distribution $p(\z)$ is usually a standard Gaussian.
The loss function in VAE encourages the model to keep its posterior distributions (encoder) close to the prior $p(\z)$, and has the form:
\vskip -0.2in
\begin{equation}\label{eq:vae_LB}
\begin{split}
 \mathcal{L}_{\text{VAE}}(\theta; \x) &= -\abbr{kl}(q_\theta(\z|\x)||p(\z)) \\ 
 &~~~~~ + \E_{q_\theta(\z|\x)}[\log p_\theta(\x|\z)] \\ 
 & \le \log p(\x)\;\;.
 \end{split}
\end{equation}
where, $KL$ is the Kullback$-$Leibler divergence $\mathrm{KL}(P\|Q) = \int_X \log\frac{dP}{dQ}\, dP$, $\E_{q_\theta(\z|\x)}[\log p_\theta(\x|\z)]$ is the reconstruction loss,
$q_\theta(\z|\x)$ is the posterior distribution approximated by the encoder, 
$p_\theta(\x|\z)$ is the posterior distribution approximated by decoder
and $\theta$ is the set of learnable parameters of the AE.

\begin{figure}[!tbp]
  \centering
  \begin{minipage}[b]{0.47\textwidth}
    \includegraphics[width=\textwidth]{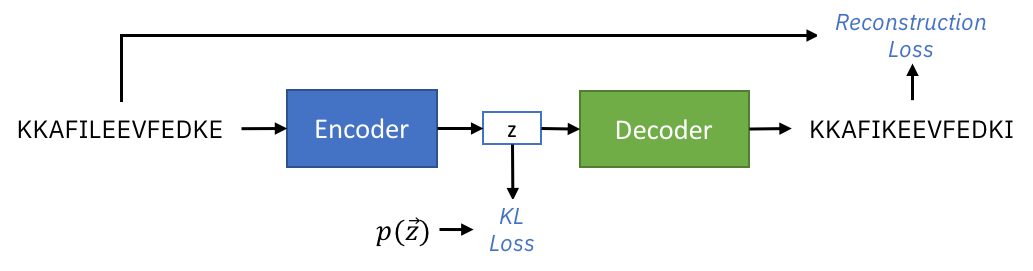}
    \caption{VAE for Modeling Sequences.}
    \label{fig:vae}
  \end{minipage}
  \hfill
  \begin{minipage}[b]{0.47\textwidth}
    \includegraphics[width=\textwidth]{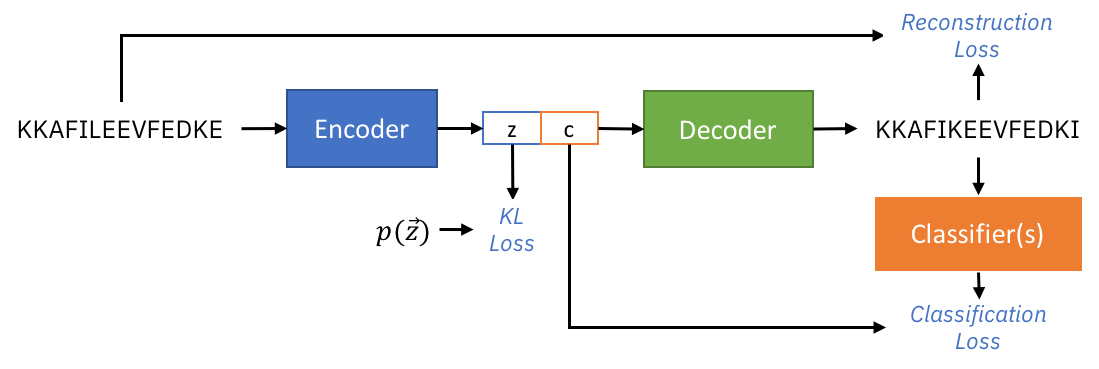}
    \caption{Controlled Sequence Generation.}
    \label{fig:cvae}
  \end{minipage}
\end{figure}

Fig. \ref{fig:vae} illustrates the VAE framework for the peptide sequence generation case.
During training,
known peptide sequences are fed into the encoder,
which generates their respective latent codes.
The latent code is later decoded into a peptide sequence using the decoder.
Next, the two components of the loss are computed and the model is updated using stochastic gradient descent (SGD).
In the figure,
we detail the inputs for computing the two components of the loss function.
After training,
one can sample a latent code $\z$ from the prior $p(\z)$ and use the decoder to generate the corresponding peptide sequence.

In our experiments,
we use a single layer Gated Recurrent Unit (GRU) RNN for both the encoder and decoder. The prior $p(\z)$ is the standard Gaussian distribution.
Previous work \cite{bowman2015generating} shows that VAE faces optimization challenges when used to model sequences. The most common issue is that the model normally sets $q_\theta(\z|\x)$ equal to the prior $p(\z)$, in this situation, the decoder essentially becomes a language model and ignores $\z$. Two solutions proposed by \citeauthor{bowman2015generating} are: (a) \emph{KL annealing}, which consists in adding a variable weight to the \emph{KL} term in the cost function at training time; and (b) \emph{word dropout}, which consists in randomly replacing a fraction of the conditioned-on tokens with a common unknown word token \emph{$<$UNK$>$}.
We adopt both solutions in our experiments.

\subsection{Semi-supervised Controlled Sequence Generation}
Although VAE is a versatile approach to train generative models, it lacks two important qualities to model AMPs:  
semi-supervised learning and controllable generation.
The set of sequences known to be AMPs is small compared to the universe of known peptides.
Therefore, when training AMP generators, 
we want to use semi-supervised methods that leverage not only the known AMP sequences but also the large number of available peptides.
An additional design requirement is
that the generative model allows ``knobs'' to control certain properties like AMP activity and toxicity.

\citeauthor{hu2017toward} (2017) proposed a VAE variant that allows both controllable generation and semi-supervised learning.
In order to perform controllable generation,
\citeauthor{hu2017toward}'s  approach augments the unstructured latent codes $\z$ with a set of structured variables $c$, which are trained to control a salient and independent attribute of the sequence: whether it is antimicrobial, toxic, soluble, etc.
Each model component of Figure \ref{fig:cvae} is alternatingly updated with a different loss function. 
For the newly introduced classifier, we minimize the loss with respect to the parameters $\theta_C$:
\begin{equation}\label{eq:cvae_LClassifier}
\begin{split}
 \LL_{\text{C}}(\theta_C) =
  &\E_{(x,c)\sim \text{lab}}[-\log q_C(c|x)] \\
+ & \E_{(x,c)\sim \text{gen}}[-\log q_C(c|x) - \beta \mathcal{H}(q_C(c'|x)]
\end{split}
\end{equation}
where the first expectation is approximated with a minibatch of the small labeled dataset,
while the second expectation is computed with a minibatch of generated data with $c,z$ sampled from the prior: $x \sim p(z) p(c) p_G(x|z,c)$.
The classifier loss requires both real and generated sequences' $c$ attribute to be classified correctly,
while minimizing the entropy $\mathcal{H}$ of the classifier encourages it to have high confidence in its predictions on generated data. For the encoder, the loss is unchanged $\LL_{\text{Enc}} = \LL_{\text{VAE}}$,
while for the decoder (Generator), the loss becomes:
\begin{equation}\label{eq:cvae_LGenerator}
\begin{split}
 \LL_{\text{Dec}}(\theta_\text{Dec}) = & \LL_{\text{VAE}} + \lambda_c \LL_c + \lambda_z \LL_z
\end{split}
\end{equation}
where the $\LL_c$ and $\LL_z$ enforce correct classification of a minibatch of ``soft'' generated sentences under the classifier and the encoder respectively.
The full expressions of $\LL_c$ and $\LL_z$ are omitted, we refer the reader to \citep{hu2017toward}.


This method gives a model with meaningful attribute codes $c$, with the major advantage that not all data needs all attribute labels $c$.
Specifically we will use a large unlabeled peptide database for the encoder and decoder losses, with a much smaller labeled dataset (peptides with reported antimicrobial annotation) for the classifier loss.
We will refer to this as semi-supervised generative modeling:
using a large unlabeled corpus to capture the distribution with VAE,
and a small labeled corpus to learn the controlling attribute code~$c$.

\section{A Dataset for Semi-supervised Training of AMP Generators}

We compiled a new two-part dataset for semi-supervised modeling of antimicrobial peptides.
The first part, \textbf{AMP-lab-15K} contains about 15K labeled peptides for which we know if they are AMPs or not.
The second part, \textbf{Uniprot-unlab-1.7M} contains just over 1.7M unlabeled sequences.
In curating AMP-lab-15K, we created the positive set by extracting experimentally validated AMP sequences from two major databases: LAMP \citep{zhao2013lamp}, and satPDB \citep{singh2015satpdb}.
LAMP is a comprehensive database of AMPs with information about their antimicrobial activity and cytotoxicity.
It consists of 3,904 natural AMPs and 1,643 synthetic peptides with antimicrobial activity.
SatPDB is an integrated database of therapeutic peptides, curated from twenty public domain peptide databases and two datasets. The duplicates between these two datasets were removed to generate a non-redundant AMP dataset. As a preprocessing step, the sequences with non-natural amino acids (B, J, O, U, X, and Z) and the ones with lower case letters were eliminated, resulting in a total of 7960 positive monomeric sequences comprised of 20 natural amino acids.
The AMP-negative peptide sequences in AMP-lab-15K are filtered out from the negative AMP dataset created by AmPEP \citep{bhadra2018ampep}. Those sequences were originally retrieved from Uniprot-Trembl comprising computer-reviewed sequences \citep{uniprot}. 
Then, sequences with any of the following annotations were removed:  AMP, membrane, toxic, secretory, defensive, antibiotic, anticancer, antiviral, and antifungal. We only considered unique sequences comprised of natural amino acids. The negative dataset contains 6948 sequences. 

For the second part, Uniprot-unlab-1.7M, the unlabeled sequences were retrieved from Uniprot-Trembl database comprising computer-reviewed sequences \citep{uniprot}.
Again duplicates and sequences with non-natural amino acids or lower case letters were removed, resulting into a total of 1.7M unlabeled sequences.

When training the VAE model, we will subselect sequences with length $\leq$ the hyperparameter \verb$max_seq_length(l)$.
Furthermore both AMP-lab-15K and Uniprot-unlab-1.7M were split into train, heldout and test set.
This reduces the available sequences for training; eg for Uniprot-unlab-1.7M the number of available training sequences are 93k for \verb$l$=25 and 168k for \verb$l$=30.

\section{Experiments}

\subsection{Experimental Setup}
The model architecture from \citep{hu2017toward} was implemented in PyTorch. For $\LL_{\text{VAE}}$ (Eq \ref{eq:cvae_LGenerator}), we use the Uniprot-unlab-1.7M unlabeled sequences, while for the first term of Eq \ref{eq:cvae_LClassifier} we use our AMP-lab-15K dataset. We consider one iteration to be a single stochastic update on both the classifier, generator, and encoder respectively with minibatch size 32. Unless otherwise noted, we pretrain the VAE for 20k iterations ($\sim 7$ epochs) followed by 5k iterations ($\sim 34$ epochs) of full model training. We noticed a slight advantage using KL annealing: from initial value $1e-5$ to $1.0$ during VAE pretraining. Further hyperparameters mostly follow \cite{hu2017toward}: learning rate = $1e-3$, balancing weights $\lambda_c = \lambda_z = \lambda_u = 0.1$, entropy $\beta=0.1$.

\begin{table}[h]
\centering
\begin{tabular}{ |c|c|c|c|c| } 
\hline
\verb$max_seq_length (l)$ & 15 & 20 & 25 & 30 \\ 
\hline
PepCVAE  & 82.17 & 82.11 & 84.33 & 83.04 \\ 
\hline
VAE  & 83.37 & 82.95 & 84.56 & 83.25 \\ 
\hline
Unlabeled & 60.16 & 47.98 & 38.26 & 28.06 \\
\hline
\end{tabular}
\caption{Accuracy of the independently trained LSTM-based AMP classifier on generated sequences and random Uniprot-Trembl sequences.}
\label{accuracy}
\end{table}

\subsection{Evaluation Metrics}
In our framework we propose 3 escalations of evaluations;  level 1 comprises preliminary automated evaluation based on peptide heuristics (sequence similarity, diversity, uniqueness, and molecular characteristics) and an external AMP classifier. Level 2 consists of an AMP potency (high/low) ranking model and \textit{ab initio} structure prediction. 
Level 3 consists of \textit{in silico} full-blown atomistic simulations and wet lab experiments. In this work we performed level 1 and 2 evaluations. More expensive level 3 evaluation will be covered in future work. 
All evaluation statistics presented in this work are averaged from 3 different runs with different random seeds. We present results with \verb$ l$ =  15, 20, 25, and 30. For evaluation we generate 5000 sequences.
\begin{figure}[h]
\centering
    \includegraphics[width=0.5\textwidth,]{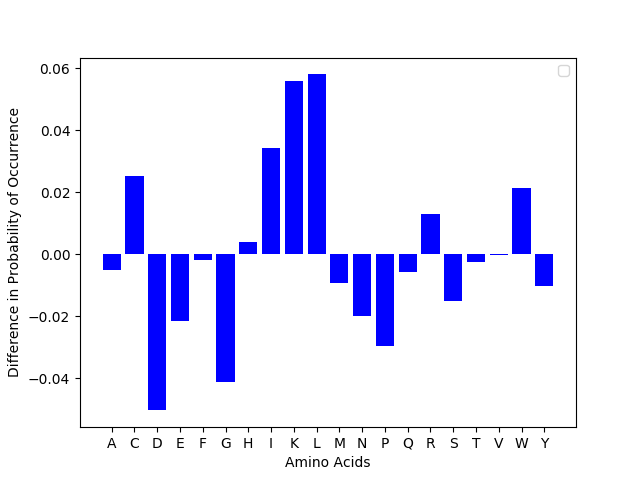}
    \caption{Difference in amino acid composition between high-confidence and low-confidence AMP sequences, as returned by the LSTM-based AMP classifier.}
    \label{fig:composition}
\end{figure}

\textbf{AMP Classification Accuracy:}
We measure the efficacy of the models on generating positive AMP sequences by assessing the accuracy using a pretrained AMP classifier for a set of generated sequences.
We independently trained a LSTM-based AMP classifier using a dataset of 8,944 labeled examples for training, 2,982 examples for validation, and another 2,982 for testing.
Our LSTM-based classifier achieves 81\% overall accuracy on the held out test examples, with 87\% accuracy for positive AMP examples and 74\% accuracy for negative examples. This  accuracy on positive samples is comparable to the models reported in literature \citep{bhadra2018ampep,veltri2018deep},
which were all trained on a much smaller set of AMPS. It should be mentioned that the relatively lower test accuracy on the negative
samples makes sense, as the majority of negative instances lack experimental validation, so there is an intrinsic low confidence associated with their label annotation. 

\textbf{Sequence Similarity and Uniqueness:} Pairwise sequence similarity was estimated using the widely used  BLOSUM62 amino acid substitution matrix \citep{henikoff1992amino}. A penalty of -10 was assigned to a gap opening and -1 penalty was assigned to a gap extension. The final results were robust against the choice of gap penalty values. The sequence similarity was normalized with the logarithm of query sequence length. A positive value suggests stronger evolutionary relationship between two sequences.

\textbf{Sequence Diversity:} Three different metrics are used to evaluate sequence diversity. (1) Language model perplexity.
A character-level LSTM language model (LM) \citep{merityRegOpt} trained on the labeled AMP/non-AMP sequences was used to 
estimate the perplexity (PPL) of the generated sequences. A lower value of PPL suggests ``closeness'' of the sequence to the original distribution it was trained on. (2) $n$-gram entropy, $E_n$,  that is the information entropy per character for $n$-grams  and is given by $E_n = -1/n \sum_{i} {p_i*lnp_i}$. Relative entropy gain, $ \delta E_n$ was defined as  $ \delta E_n = (E_n^{cvae} - E_n^{orig}) / (E_n^{vae} - E_n^{orig})$. We further mixed generated samples with original samples at a 1:1 ratio and again estimated relative entropy gain in the above-mentioned manner,  which is captured in $ \delta E_n^{mix}$. (3) The diversity is also estimated by measuring the number of shared $n$-grams \citep{osmanbeyoglu2011n} for different values of $n$ between the generated sequences and the original ones, which we refer as $S_n$. Therefore, a value of $S_n^{cvae} / S_n^{vae} <1$ implies more diversity of PepCVAE sequences at a particular $n$ compared to that of the VAE ones.   

\textbf{Molecular Characteristics:}
Peptide characteristics, \textit{e.g.} hydrophobicity, charge, were estimated using the GlobalAnalysis method in modLAMP \citep{muller2017modlamp}. 

\textbf{AMP Raking Model:}
In level 2 evaluation, PepCVAE-generated sequences with $> 0.98$ antimicrobial probability, as predicted by the external LSTM classifier, were selected and ranked according to their predicted potency (high/low). For ranking, an LSTM model was trained on 1200 independent sequences with broad-spectrum antimicrobial activities from \citep{pirtskhalava2015dbaasp} and yielded a test accuracy of $ 70\%$. 

\textbf{Ab Initio Structure Prediction:} 
For structure prediction of  sequences, PEP-FOLD3 server \citep{lamiable2016pep} was used, which employs structural alphabets (SA) to describe the structure of four consecutive amino acids, couples the predicted series of SA letters to a greedy algorithm and a coarse-grained force field, generates 3D structures and finally sorts them according to energy.


\subsection{Experimental Results}


Table \ref{accuracy} presents the classification accuracy of the AMP sequences generated by PepCVAE. Specifically, we generate sequences given attribute code c that matches AMP, and then use the pre-trained LSTM-based AMP classifier to assign AMP/non-AMP labels to the generated sequences. We compare the classification accuracy of PepCVAE-generated AMPs  with the one obtained by using a plain vanilla VAE trained solely on AMP sequences. The probability of randomly selected unlabeled sequences from Uniprot-Trembl database to be predicted as AMP by the classifier is also shown for comparison. It is evident from Table \ref{accuracy}, that the generated AMP sequences by both PepCVAE and plain VAE are predicted to be ``active'' with a probability $>82\%$,
which is significantly higher than the predicted probability for the unlabeled training sequences of all lengths.


In Fig. \ref{fig:composition},
we show, for a set of 5K sequences generated by PepCVAE, 
the unigram distribution difference between the high confidence and low confidence samples, as returned by the LSTM-based AMP classifier.
This difference expresses the positive contribution of cationic  (K, R) and hydrophobic (A, C, L I, W) amino acids towards determining AMP character, which is consistent with features identified by existing AMP classifiers \citep{bhadra2018ampep,veltri2018deep}.   



\begin{figure}[!tbp]
  \centering
  \begin{minipage}[b]{0.45\textwidth}
    \includegraphics[width=\textwidth]{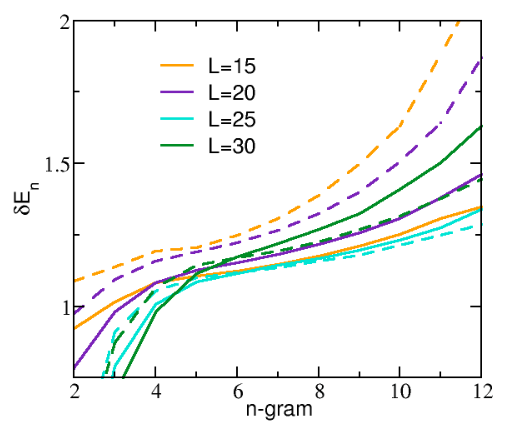}
    \caption{Relative entropy gain of PepCVAE sequences w.r.t VAE (Solid lines - without mixing, dashed lines - with mixing). See Evaluation Metrics  section for details.}
    \label{fig:entropy}
  \end{minipage}
  \hfill
  \begin{minipage}[b]{0.45\textwidth}
    \includegraphics[width=\textwidth]{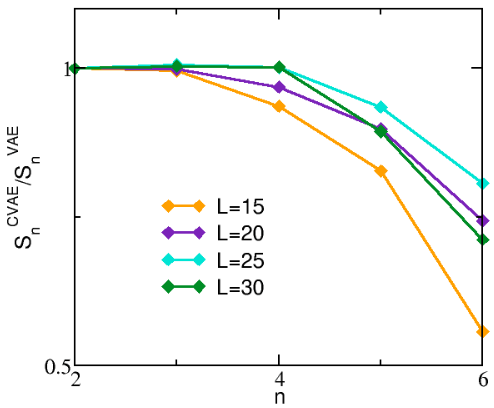}
    \caption{n-gram shared similarity, $S_n$ of PepCVAE over VAE sequences as a function n.  $S_n < 1$ implies higher diversity of PepCVAE sequences.}
    \label{fig:ngram}
  \end{minipage}
\end{figure}

\subsubsection{Peptide Heuristics}
\begin{table*}[h] 
\centering
\begin{tabular}{ |c|c|c|c|c|c|c| } 
 \hline
  Model & Length & Uniq-3 & Uniq-4 &  S\textsuperscript{self} & S\textsuperscript{orig}   & PPL \\ 
 \hline
 Training & 15.53  & 16.00 & 52.40  & -6.49 & NA & 3.57 \\ 
 \hline
 PepCVAE & 16.55 & 11.40 & 68.10   &  -6.52 & -6.62 & 30.27  \\ 
 \hline
 VAE & 15.12 & 9.90 & 72.00 & -7.49 & -7.51   & 32.72 \\
 \hline
 Unlabeled & 18.84 & 4.90 & 68.70   & -5.68 & -6.71  & 28.77  \\
 \hline
 Random & 15.03  & 0.02 & 83.00   & -6.90 & -7.63& 38.10 \\
 \hline
\end{tabular}
\caption{Sequence heuristics for $ l $ = 25.
Positive instances from  PepCVAE  and plain VAE were compared with training AMPs, random sequences, and unlabeled Uniprot sequences.}
\label{heuristics}
\end{table*}
Given that both cVAE and VAE generate AMP sequences with high probability, next we estimate a number of heuristics that give some clues about how similar/dissimilar the generated AMPs are compared to the ones present in the training set. 
The heuristics are sequence length, fraction of unique 3 and 4-gram, and sequence similarity estimated by using standard amino acid substitution matrix, and language model perplexity (PPL). 
Table \ref{heuristics} presents these estimates for \verb$l$=25. 
For the purpose of comparison, the values corresponding to random peptides and unlabeled sequences from Uniprot of similar length are also provided. The VAE sequences  are closer in length to the original ones, while the PepCVAE ones are relatively longer. Both methods result into AMPs with lower uniqueness at a 3-gram level and higher uniqueness at a 4-gram level, with respect to the  original ones. PepCVAE sequences appear closer to the original ones in terms of 3 and 4-gram uniqueness. 

The average pairwise sequence similarity within the set of generated sequences itself, S\textsuperscript{self}, is consistently lower for PepCVAE compared to VAE (Table \ref{heuristics}). S\textsuperscript{self} of cVAE sequences is closer to that of the biological distribution (both training and unlabeled). This is meaningful, as a higher self-similarity or ``homology'' implies stronger evolutionary relation between the sequences. In this sense, the extent of homology within the PepCVAE sequences is higher than the VAE ones and matches more closely to that of existing AMPs. S\textsuperscript{orig} further suggests that PepCVAE sequences possess stronger evolutionary relationship with the actual AMPs as well as the unlabeled biological peptides. The high evolutionary dissimilarity of VAE sequences (both with self and with biological) is reminiscent of random sequences (refer Table \ref{heuristics}). It is likely that PepCVAE learns a more ``biologically plausible'' latent space by exploiting a much larger dataset, that includes unlabeled and negative  sequences as well as the positive ones. These results suggest that PepCVAE architecture intrinsically inserts more ``biological'' character/context  during controlled generation of AMPs.  
The perplexity value (PPL) returned by a LSTM language model that was independently trained on the AMP-lab-15K dataset further confirms this observation. PepCVAE sequences are low in perplexity and are closer to biological sequences, whereas high PPL of VAE samples implies more random-like character.


\subsubsection{Sequence diversity}
Next, we analyze the diversity of the generated sequences in terms of $n$-gram entropy gain (see Evaluation Metrics section). Figure \ref{fig:entropy} plots the relative entropy gain of PepCVAE sequences with respect to VAE, $ \delta E_n$, as a function of $n$-gram size. We observe that $\delta E_n \leq 1$ for $n \leq 3$, while increasing to values  $> 1$ for larger $n$. 
This result implies that, although VAE sequences are more diverse locally ($n <4$), PepCVAE sequences demonstrate strong long-range diversity. Consistent with this result, the $n$-gram similarity, \textit{i.e.} fraction of shared $n$-grams with training AMPs, is lower for PepCVAE with respect to VAE for $n \geq 3$ (Figure \ref{fig:ngram} and Evaluation Metrics  section). Even though PepCVAE generates diverse sequences, at short range it is still  consistent with biological sequences, as evident from the language model perplexity values (Table \ref{heuristics}). In summary, the PepCVAE sequences show stronger diversity at higher $n$-grams. High peptide diversity compared to existing AMPs is a desired feature, while designing next-generation antimicrobials, as that can potentially help prevent antimicrobial resistance. 

\begin{figure*}[h]
\centering
\includegraphics[width=0.9\textwidth]{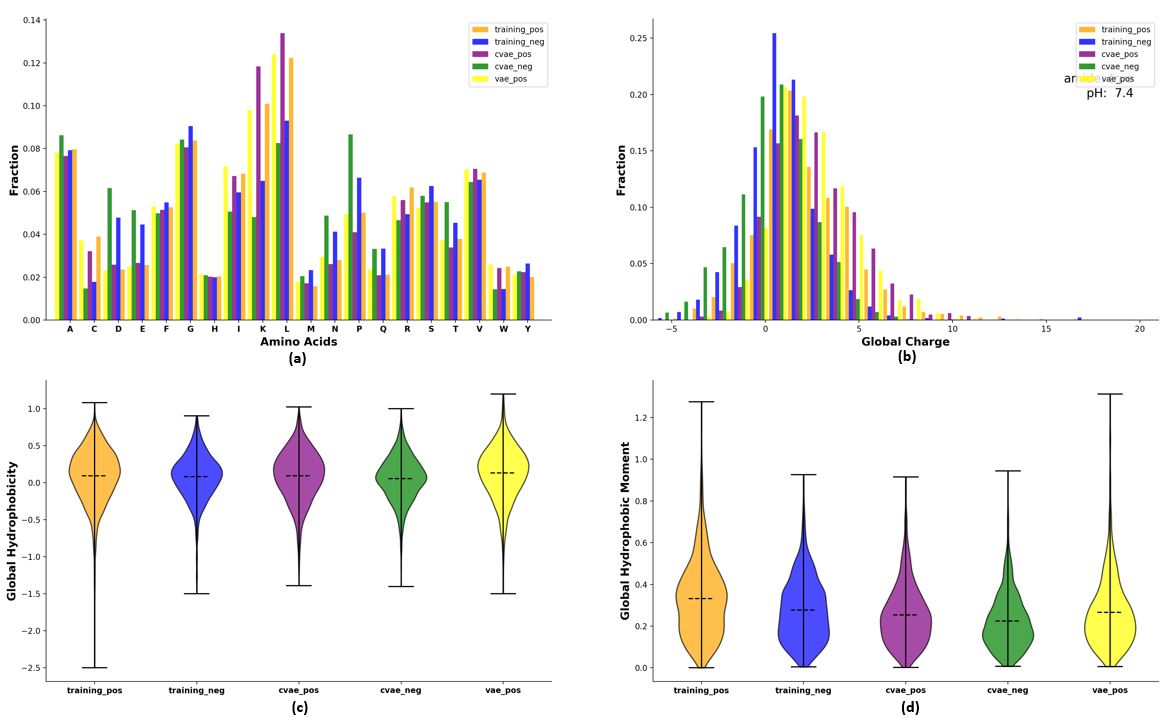}
\caption{Comparison of molecular characteristics for  $l \eqslantless 25$ between training data (training-pos - orange, training-neg - blue), PepCVAE sequences (cvae-pos - purple, cvae-neg - green), VAE sequences (vae-pos - yellow). Horizontal dashed lines account for the mean. Whiskers extend to the most extreme non-outlier data points. (a) amino acid distribution, (b) total charge distribution, (c) Eisenberg hydrophobicity, and (d) Eisenberg hydrophobic moment.}
\label{fig:peptide}
\end{figure*}


\begin{table*}[h]
\centering
\begin{tabular}{ |c|c|c|c|c|c| } 
 \hline
 Sequence & $ l $ & Charge  & H & $\mu$H & Structure\\ 
 \hline
 MWHFIWYLILLPRR & 13 & 3.0 & 0.30 & 0.37 & Helix\\
 \hline
 LWNYWFLWSAFRAF & 14  & 2.0 & 0.45 & 0.19& Helix\\
 \hline
 YHSIFFCFKKIKAK & 14 & 5.0 & 0.07 & 0.28 & Helix \\
 \hline
 IIYLIWWWLNWV & 12 & 1.0 & 0.85 & 0.39 & Helix\\
 \hline
 HKERRWRYW  & 9 & 4.0 & -0.92 & 0.35 & Helix\\
\hline
\end{tabular}
\caption{High-potency AMP sequences (and their features)  generated by PepCVAE with $l$ =15.}
\label{peptide}
\end{table*}
\subsubsection{Molecular Characteristics}
Figure \ref{fig:peptide} compares PepCVAE and VAE sequences with the training data in terms of molecular features, \textit{e.g.} charge, hydrophobicity (H), and hydrophobic moment ($\mu$H), which are of particular interest, as they  play a key role in determining the membrane binding specificity \citep{fjell2012designing}. 

The amino acid composition, net positive charge, and hydrophobicity (H) (Fig. \ref{fig:peptide}a-c) of generated AMPs by  PepCVAE and VAE matches well to the training data, suggesting both models perform equally well in capturing the charge patterning, hydrophobicity, and composition within AMPs.  
The hydrophobic moment ($\mu$H), a qualitative measure of the helical character within the sequence, is another frequently used descriptor used in cationic amphiphathic  AMP classification \citep{bhadra2018ampep}. Both VAE and PepCVAE generate sequences with hydrophobic moment values consistent with, albeit slightly lower than,  existing AMPs. 

\subsubsection{AMP Potency ranking and structure prediction}
We first selected 45  high probability AMPs out of 5000 PepCVAE sequences and ranked them according to the predicted potency. Next, 3D models of the final 11 high-potency AMP sequences were constructed using the PEP-FOLD3 server \citep{lamiable2016pep}. 
Out of those 11 candidates, the lowest energy model of 9 sequences consistently exhibited a helix (for examples see Table \ref{peptide}), one showed an extended structure, and one revealed coil.  As amphipathic helices are abundant in antimicrobial peptides and determine their activity, this multi-level \textit{in silico} screening scheme successfully identifies high potency, broad-spectrum antimicrobial candidates.

\section{Conclusion and Future Work}
\noindent
We present a peptide sequence design  framework PepCVAE based on a semi-supervised variational autoencoder model for generating novel antimicrobial peptide molecules. We curated a dataset that comprises a large number (1.7M) of unlabeled peptide sequences and a smaller set (15k) of labeled (AMP/non-AMP) sequences.
The model architecture allows learning of a representation where desired properties are disentangled,
and so can handle controlled generation of peptide sequences with AMP/non-AMP characteristics.
Extensive analysis of the generated antimicrobial sequences reveals that the proposed framework is capable of learning and generating from a richer representation and yields AMPs that are closer to the original distribution, when compared with a vanilla VAE trained solely on AMP sequences.
The  generated AMPs from our architecture exhibit high diversity, particularly at a higher $n$-gram level, while still retaining biological characteristics, such as stronger homology and high helicity.
These peptide characteristics suggest that the present framework is well-suited for therapeutic molecule design, where it is important to  maintain control over ``knobs'' or attributes, while generating novel samples. 
In future, we plan to validate the generated AMP sequences using \textit{in silico} modeling and wet lab experiments.



\bibliographystyle{nips_2018}
\bibliography{nips_2018}

\end{document}